\documentclass[a4paper]{article}
\usepackage{RR}
\RRNo{6214}

\usepackage[english]{babel}

\usepackage{color}

\usepackage[cmex10]{amsmath}
\usepackage{amsfonts}
\usepackage{amsthm}
\newtheorem{theorem}{Theorem}
\newtheorem{lemma}[theorem]{Lemma}

\newtheorem{definition}[theorem]{Definition}

\newtheorem{examp}{{Example}}
\newenvironment{example}[1]
{\begin{examp}[{#1}]\rm}
{\eitemproof\end{examp}}

\newcommand{\elim}[2]{\left[#1\right]_{#2}}
\newcommand{\bigelim}[2]{\bigl[#1\bigr]_{#2}}
\newcommand{\fusion}[2]{[\![#1]\!]_{#2}}

\newcommand{\para}{\;\|\;}

\newcommand{\Ra}{\Rightarrow}

\newcommand{\eitemproof}{\hfill$\diamond$}
\newcommand{\behav}{E}

\newcommand{\controlled}{{\bf c}}

\newcommand{\uncontrolled}{{\bf u}}

\newcommand{\vis}{{\bf vis}}
\newcommand{\loc}{{\bf loc}}
\newcommand{\profile}{\pi}

\newcommand{\true}{{\tt T}}
\newcommand{\false}{{\tt F}}

\newcommand{\beq}{\begin{eqnarray}}
\newcommand{\eeq}{\end{eqnarray}}
\newcommand{\beqq}{\begin{eqnarray*}}
\newcommand{\eeqq}{\end{eqnarray*}}
\newcommand{\bea}{\begin{array}}
\newcommand{\eea}{\end{array}}

\newcommand{\intersect}{\cap}
\newcommand{\union}{\cup}

\newcommand{\product}{\mathbin{||}}

\newcommand{\richcomp}{{\it RC}}
\newcommand{\rcname}{X}
\newcommand{\setsofcontracts}{\left\{\cgte\right\}}
\newcommand{\optimplementation}{\left[\comp\right]}

\newcommand{\runs}{\mathcal{R}}

\newcommand{\assp}{A}
\newcommand{\prom}{G}

\newcommand{\cgte}{C}
\newcommand{\comp}{M}

\newcommand{\complement}[1]{\neg{#1}}
\newcommand{\join}{\sqcup}
\newcommand{\meet}{\sqcap}
\newcommand{\bigmeet}{\mbox{\Large $\meet$}}

\newcommand{\maxcomp}[1]{{\comp_{#1}}}

\newcommand{\compref}{\preceq}

\newcommand{\cgref}{\preceq}

\newcommand{\contract}{contract}
\newcommand{\Contract}{Contract}
\newcommand{\contracts}{contracts}

\newcommand{\pprime}{'}

\def\IEEEproof{\textbf{Proof:}~}
\def\endIEEEproof{\textbf{\hfil$\Box$}}

\begin{document}

\RRdate{June 2007}

\RRetitle{A Generic Model of Contracts for Embedded Systems\thanks{
This research has been developed in the framework of the European 
SPEEDS integrated project number 033471.}}

\titlehead{A Generic Model of Contracts for Embedded Systems}

\RRtitle{Un modele g\'en\'erique de contrats pour les syst\`emes
embarqu\'es}

\RRauthor{Albert Benveniste\thanks{IRISA / INRIA, Rennes, France,
albert.benveniste@irisa.fr}, Beno\^{\i}t Caillaud\thanks{IRISA /
INRIA, Rennes, France, benoit.caillaud@irisa.fr}, Roberto
Passerone\thanks{University of Trento \& PARADES EEIG, Trento \& Rome,
Italy, roberto.passerone@unitn.it}}

\authorhead{Benveniste, Caillaud \& Passerone}

\RRabstract{
We present the mathematical foundations of the contract-based model
developed in the framework of the SPEEDS project.
SPEEDS aims at developing methods and tools to support ``speculative
design'', a design methodology in which distributed designers develop
different aspects of the overall system, in a concurrent but controlled way.
Our generic mathematical model of contract supports this style of
development.
This is achieved by focusing on behaviors, by supporting the notion of
``rich component'' where diverse (functional and non-functional) aspects of
the system can be considered and combined, by representing rich components
via their set of associated contracts, and by formalizing the whole process
of component composition.
}

\RRresume{ Ce rapport pr\'esente les fondements math\'ematiques du
mod\`ele de contrats con\c{c}u dans le cadre du projet
SPEEDS. L'objectif du projet SPEEDS est de d\'evelopper les outils et
les m\'ethodes supportant un ``processus de conception sp\'culatif'',
dans lequel diff\'erentes \'equipes de conception peuvent contribuer
\`a la conception d'un syst\`eme de f\c{c}on concurrente, mais
n\'eanmoins controll\'ee. Le mod\`ele de contrats concerne les
comportements du syst\`eme projet\'e et permet une mod\'elisation de
celui-ci par assemblage de ``composants riches'', dont les
diff\'erents aspects comportementaux sont d\'ecrits pas des ensembles
contrats, regroup\'es par ``points de vues''.  }

\RRkeyword{system design, component based design, contract based design, assume-guarantee reasoning}

\RRmotcle{concepton syst\`eme, conception par composants, conception par contrats, raisonnement hypioth\`ese-garantie}

\RRprojet{S4}

\RRtheme{\THCom}

\URRennes

\makeRR

\section{Introduction} \label{sec:intro}


Several industrial sectors involving complex embedded systems design have
recently experienced drastic moves in their organization---aerospace
and automotive being typical examples. 
Initially organized around large, vertically integrated companies supporting
most of the design in house, these sectors were restructured in the 80's 
due to the
emergence of sizeable competitive suppliers.
OEMs performed system design and
integration by importing entire subsystems from suppliers.
This, however, shifted a significant portion of the value to the suppliers,
and eventually contributed to late errors
that caused delays and excessive additional cost during the system
integration phase.

In the last decade, these industrial sectors went through a profound
reorganization in an attempt by OEMs to recover value from the supply chain,
by focusing on those parts of the design at
the core of their competitive advantage.  The rest of the system was
instead centered around standard platforms that could be developed and
shared by otherwise competitors.  Examples of this trend are AUTOSAR
in the automotive industry~\cite{Damm06}, and 
Integrated Modular Avionics (IMA) in aerospace~\cite{Butz07}.  This new
organization requires extensive virtual prototyping and design space
exploration, where component or subsystem specification and
integration occur at different phases of the design, including at the early
ones~\cite{ASV2007}.

Component based development has emerged as the technology of choice to
address the challenges that result from this paradigm shift.
In the particular context of (safety
critical) embedded systems with complex OEM/supplier chains, the
following distinguishing features must be addressed. First, the need
for high quality, zero defect, software systems calls for techniques in
which component specification and integration is supported by clean
mathematics that encompasse both static and \emph{dynamic} semantics---this
means that the behavior of components and their
composition, and not just their port and type interface, must be
mathematically defined. Second, system design includes various
aspects---functional, timeliness, safety and fault tolerance, 
etc.---involving different teams with different skills using
heterogeneous techniques and tools. Third, since the structure of the
supply chain is highly distributed, a precise separation of
responsibilities between its different actors
must be ensured. This is addressed by relying on {contracts}.
Following \cite{Damm05} a contract is a component model 
that sets forth the {assumptions} under which the component
may be used by its environment, and the corresponding {promises} that
are guaranteed under such correct use.

The semantic foundations that we present in this paper are designed to
support this methodology by addressing the above three issues.  At
its basis, the model is a language-based abstraction where composition
is by intersection.  This basic model can then be instantiated to
cover functional, timeliness, safety, and dependability requirements
performed across all system design levels.  No particular model of
computation and communication is enforced, and 
continuous time dynamics such as those needed in physical system
modeling is supported as well.
On top of the basic model, we build the notion of a contract, which is
central to our methodology, by distinguishing between assumptions and
promises.
This paper focuses on developing a generic 
compositional theory of contracts, providing 
relations of contract satisfaction and refinement called
dominance, and the derivation of operators for the correct construction of
complete systems.
In addition to traditional parallel composition, and to enable formal
multi-viewpoint analysis, our model includes boolean meet and join operators
that compute conjunction and disjunction of contracts.
We also introduce a new operator, called fusion, that combines composition
and conjunction to compute the least specific contract that satisfies a set
of specifications, while at the same time taking their interaction into
account.

The paper is organized as follows.  The principles of our approach are
presented in Section~\ref{sec:hrc}.  Contracts and implementations are
introduced in Section~\ref{fguioege} and corresponding operations are
studied in Section~\ref{eprfouiehpuio}.  The concept of rich component
is formalized in Section~\ref{epfguioioe}, by introducing the
contracts attached to it.  In Section~\ref{fuohepuo} we formalize the
concept of designer responsibilities through the notion of
controlled/uncontrolled port and we refine our theory of contracts
accordingly.  How we encompass the different viewpoints is sketched in
Section~\ref{er;ofuierhopuif} and related work is discussed in
Section~\ref{sec:related}.

\label{sec:contracts}
\newcommand{\pre}{{\it pre}}
\newcommand{\tle}{{\rm TLE}}
\newcommand{\never}{{\it never}}
\newcommand{\always}{{\it always}}

\section{Principles of Assume/Guarantee Reasoning}
\label{sec:hrc}
The main element of our semantic model is a \emph{Heterogeneous Rich
Component}, or simply a component.
A component consists of an \emph{interface}, its \emph{expected
behavior}, and, optionally, one or more \emph{implementations}.
The interface is a set of ports and flows,
used by the component to communicate with the rest of the system and with
the environment.
The expected behavior is described by one or several
\emph{assumption}/\emph{promise} pairs, called \emph{contracts}.
Contracts can be combined together using three composition operators:
greatest lower bound, parallel composition and fusion.
The greatest lower bound is used to compose contracts referring to the same
component and which use only variables and flows visible from the
environment.
Parallel composition is used to compute the contract resulting from the
composition of several components.
Fusion generalizes these two operators, and is capable of handling all
cases.
In particular, it is used to compose contracts whenever the greatest lower
bound and parallel composition operators are inappropriate, for instance
when contracts share local variables or flows.
Thus, fusion is the implicit composition of contracts, whenever more than
one contract is attached to a component.
Implementations may be attached to a component, and are usually expressed as
extended state machines, or as host tool models.
We define several relations between components, contracts and
implementations.
\begin{itemize}
  \item The \emph{compatibility} relation relates sets of components.
A set of
    components are \emph{incompatible} whenever for all environments, at
    least one of the assumption of at least one component is violated.

  \item Contract \emph{dominance} relates assumptions and promises of two
    contracts.  A contract dominates another when it has weaker assumptions
    and stronger promises.

  \item \emph{Satisfaction} relates implementations to contracts.
    An implementation satisfies a contract whenever its behavior, modulo the
    assumptions, are consistent with the promises.
  \item \emph{Refinement} relates implementations.
    An implementation refines another whenever it has fewer behaviors.
\end{itemize}
Throughout this paper we shall need an abstract notion of ``assertion''.
The only facts we need to know about assertions can be 
summarized as follows:
\begin{itemize}
\item Each assertion $\behav$ possesses a set of \emph{ports} and a set of
  \emph{variables} that are the vehicle for interaction.
\item An assertion is identified with the set of runs it accepts.
  A run assigns a history to each variable and port of the assertion.
  We assume that a proper notion of ``complement'' for an assertion $\behav$
  is available, denoted by $\complement{\behav}$.
\item When seen as sets of runs, assertions compose by intersection---note
  that such an operation is monotonic w.r.t.\ inclusion of sets of runs.
  When performing this composition, we assume that the
  appropriate inverse projections have been performed to equalize the
  sets of ports and variables.
  Products are equivalently denoted by $\behav_1 \times \behav_2$ or
  $\behav_1 \cap \behav_2$.
\end{itemize}

\section{Rich Components, Contracts, Implementations}
\label {fguioege}
\begin{definition}[Implementation] \label {erpwifuerilu}
An \emph{implementation} is simply an assertion, that is, a set of runs.
\end{definition}
We denote implementations by the symbol $\comp$ (for ``machine'').
Implementations are
ordered according to the runs they contain.  An implementation $\comp$
\emph{refines} an implementation $\comp'$, written $\comp \compref
\comp'$ if and only if $\comp$ and $\comp'$ are defined over the same
set of ports and variables, and
\[
\comp \subseteq \comp'.
\]
Products preserve implementation refinement.

A \emph{\contract} says that under certain assumptions, behaviors are
guaranteed to be confined within a certain set.

\begin{definition}[{\Contract}]\label{def:contract}
A \emph{\contract} $\cgte$ is a pair $(\assp, \prom)$, where $\assp$,  the
\emph{assumption,} and $\prom$, the \emph{promise,} are assertions over the
same alphabet.
\end{definition}
Whenever convenient, we shall denote the assumption and promise of contract
$\cgte$ by $\assp_\cgte$ and $\prom_\cgte$.
The interpretation of a {\contract} is made precise by the following
definition.
\begin{definition}[Satisfaction]\label{def:satisfaction}
An implementation $\comp$ satisfies a {\contract} $\cgte = (\assp, \prom)$, written
$\comp \models \cgte$, if and only if
\[
\comp \intersect \assp \subseteq \prom.
\]
\end{definition}
Satisfaction can be checked using the following equivalent formulas, where
$\complement{\assp}$ denotes the set of all runs that are not runs of $\assp$:
\begin{eqnarray*}
\comp \models \cgte \ \iff \ \comp \subseteq \prom \union \complement{\assp}
\ \iff \ \comp \intersect (\assp \intersect \complement{\prom}) = \emptyset
\label{th:alt_satisfaction}
\end{eqnarray*}
There exists a unique maximal implementation satisfying a contract 
$\cgte$, namely: 
\beq
\maxcomp{\cgte} &=& G \cup \complement{\assp}
\label {'eotj[i}
\eeq
\begin{definition}[Rich Component] \label {epfiueipu}
A \emph{rich component} is a tuple 
\beq
\richcomp &=& \bigl(\rcname, \setsofcontracts, \optimplementation\bigr)
\label {epruifgeipu}
\eeq
In (\ref{epruifgeipu}), $\rcname$ is the \emph{name} of the rich
component, $\setsofcontracts$ is a (possibly empty) set of contracts,
and $\optimplementation$ is an (optional) implementation such that
$\optimplementation\models\setsofcontracts$, where the meaning of the
latter property is postponed to Definition \ref{pweifugbepruif}.
\end{definition}
%

\paragraph*{Canonical forms}
Note that $\maxcomp{\cgte}$ is to be interpreted as the implication
$\assp\Rightarrow\prom$. We have that $\comp\models(\assp,\prom)$, if
and only if $\comp\models(\assp,\maxcomp{\cgte})$, if and only if
$\comp\subseteq\maxcomp{\cgte}$.  Say that contract $\cgte = (\assp,
\prom)$ is in \emph{canonical form} when $\prom = \maxcomp{\cgte}$,
or, equivalently, when $\complement{\assp}\subseteq\prom$
or  when $\complement{\prom}\subseteq\assp$. Thus, every
{\contract} has an equivalent {\contract} in canonical form, which can
be obtained by replacing $\prom$ with $\maxcomp{\cgte}$.  Hence,
working with {\contracts} in canonical form does not limit
expressiveness.  The operation of computing the canonical form is well
defined, since the maximal implementation is unique, and it is
idempotent.
\beq\bea{l}
\mbox{\emph{In the following, we assume that}}
\\
\mbox{\emph{all {\contracts} are in canonical form.}}
\eea
\label {erotihjpui}
\eeq
%

This assumption serves two purposes: (i) To have a unique
representation of contracts, considered up to equivalence. (ii) To
simplify the definition of contract composition operators.

\begin{example}{Running example: control/monitoring unit}
\label {rfguioerhfui}
Throughout this paper, we develop the system of Figure~\ref{erpfiuerhpifu}
to illustrate our notion of contract and its use. It consists
of a control unit interacting with a monitoring unit.
The system is subject to two independent faults, $f_1$ for
the control unit, and $f_2$ for the monitoring unit.
\begin{figure}[htb]
\centerline{\input{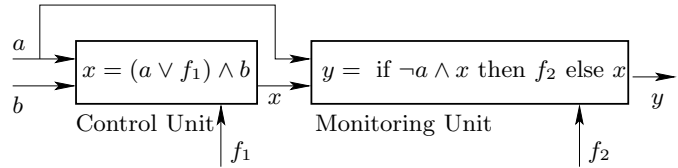}}
\caption[kuyf]{\small Running example: control/monitoring system.}
\label {erpfiuerhpifu}
\end{figure}
The nominal behavior of the system (when $f_1=\false$) is that
it should deliver $y=a\wedge b$ at its output. When safe
($f_2=\false$), the monitoring unit ensures that, if the control unit
gets faulty ($f_1=\true$), the overall system is shut down
($y=\false$) unless $a=\true$ holds. Thus the overall system requirement is
to maintain the Top Level Exception
$\tle=\neg a \wedge y$ false.  This TLE may, however, get
violated if the monitoring unit gets faulty too ($f_2=\true$).
These requirements are summarized by the two contracts
$\cgte$, for the nominal mode, and $\cgte'$ for the exception mode:
\beq\bea{rclcl}
\cgte &=& (\,\neg f_1\,,\,y=a\wedge b\,) &:&\mbox{nominal mode}
\\
\cgte' &=& (\,\neg f_2\,,\,\neg\tle\,) &:&\mbox{exception mode}
\eea
\label {rpefuiehpu}
\eeq
This separation of concerns into nominal and exception mode is 
similar to the separation of viewpoints (functional,
timed, safety, etc) when handling components via their contracts.
\end{example}

\section{Operations on contracts}
\label {eprfouiehpuio}
\subsubsection{\textbf{Boolean algebra}}
As usual, it will be extremely useful to have an algebra of contracts,
providing ways of expressing more complex contracts from simpler
ones. The following relation of dominance formalizes substituability
for contracts and induces a boolean algebra of contracts, which
provides such a logic of contracts.

\begin{definition}[Dominance]\label{def:dominance_restricted}
Say that {\contract} $\cgte = (\assp, \prom)$ dominates 
{\contract} \mbox{$\cgte' = (\assp', \prom')$,} written $\cgte \cgref
\cgte'$, if and only if $\assp \supseteq \assp'$ and $\prom \subseteq
\prom'$.
\end{definition}
Dominance amounts to relaxing assumptions and reinforcing
promises. Note that $\cgte \cgref \cgte'$ and $\cgte' \cgref \cgte$
together imply $\cgte = \cgte'$. Hence, dominance is a partial order
relation. Furthermore,
\beq
\cgte \cgref \cgte' &\implies& \maxcomp{\cgte} \models \cgte'
\label {erpotgihpeori}
\eeq
but the converse is not true. Property (\ref{erpotgihpeori}) implies
the following result:
\begin{lemma} \label {erpifugefipu}
If $\comp\models\cgte$ and $\cgte\cgref\cgte'$, then $\comp\models\cgte'$.
\end{lemma}
The following theorem defines the boolean algebra over contracts,
implied by $\cgref$. Its proof is straightforward and left to the
reader.

\begin{theorem} [Boolean algebra of contracts] 
\label {th:bounds_restr}
Let $\cgte_1 = (\assp_1, \prom_1)$ and $\cgte_2 = (\assp_2, \prom_2)$
be {\contracts}.  Then, the greatest lower bound of $\cgte_1$ and
$\cgte_2$, written $\cgte = \cgte_1 \meet \cgte_2$, is given by $\cgte
= (\assp, \prom)$ where $\assp = \assp_1 \union \assp_2$ and $\prom =
\prom_1 \intersect \prom_2$. Note that the so defined pair
$(\assp,\prom)$ is in canonical form.

Similarly, the least upper bound of $\cgte_1$ and $\cgte_2$, written
$\cgte = \cgte_1 \join \cgte_2$, is given by $\cgte' = (\assp',
\prom')$ where $\assp = \assp_1 \intersect \assp_2 $ and $\prom =
\prom_1 \union \prom_2 $. Note that the so defined pair
$(\assp,\prom)$ is in canonical form.

The minimal and maximal contracts are $\bot=(\runs, \emptyset)$ and
$\top=(\emptyset, \runs)$, respectively, where $\runs$ denotes the set of
all runs.

Finally, the complement of $\cgte$ is the {\contract}
$\complement{\cgte}$ such that $\complement{\cgte} =
(\complement{\assp}, \complement{\prom}) $; it satisfies 
$\complement{\cgte}\meet\cgte=\bot$ and 
$\complement{\cgte}\join\cgte=\top$.
\end{theorem} 

\begin{example}{Running example: greatest lower bound}
\label {e;rouifhe}
The two contracts of (\ref{rpefuiehpu}) represent two viewpoints
attached to a same component, corresponding to the nominal and
exception modes, respectively. These two contracts involve the same
set of ports.  Combining them is by computing their greatest lower
bound.  Putting these two contracts in canonical form and then taking
their greatest lower bound yields:
%
\beqq
\cgte \;\meet\; \cgte'
= \left(\,\neg (f_1\wedge f_2)\,,\,
\left[\bea{cl}
& \neg f_1 \Ra y=a\wedge b
\\
\wedge & \\
& \neg f_2 \Ra \neg 
\tle\eea\right]\,\right)
\eeqq
This contract assumes that no double failure occurs. Its promise is
the conjunction of the promises of $\cgte$ and $\cgte'$.  Expanding
the promise of this global contract leads to a cumbersome formula,
hardly understandable to the user, so we discard it.
\end{example}

\subsubsection{\textbf{Parallel composition}}
Contract composition formalizes how contracts attached to different
rich components should be combined to represent a single, compound,
rich component.
Let $\cgte_1 = (\assp_1, \prom_1)$ and $\cgte_2 = (\assp_2, \prom_2)$ be
{\contracts}.
First, composing these two contracts amounts to composing their promises.
Regarding assumptions, however, the situation is more subtle.
Suppose first that the two contracts possess disjoint sets of ports and
variables.
%
Intuitively, the assumptions of the composite should
be simply the conjunction of the assumptions of the rich components,
since the environment should satisfy all the assumptions.
In general, however, part of the assumptions $\assp_1$ will be already
satisfied by composing $\cgte_1$ with $\cgte_2$, acting as a partial
environment for $\cgte_1$.
Therefore, $\prom_2$ can contribute to relaxing the assumptions $\assp_1$.
And vice-versa.
Whence the following definition:

\begin{definition}[Parallel composition of contracts]
\label{def:composition_restr}
Let $\cgte_1 = (\assp_1, \prom_1)$ and $\cgte_2 = (\assp_2, \prom_2)$
be {\contracts}. Define $\cgte_1 \product \cgte_2$ to be the contract
$\cgte = (\assp, \prom)$ such that:
\begin{eqnarray*}
\assp & = & (\assp_1 \intersect \assp_2) \union \complement{(\prom_1
\intersect \prom_2)}, \\
\prom & = & \prom_1 \intersect \prom_2.
\end{eqnarray*}
Note that the so defined contract is in canonical form.
\end{definition}
The following result expresses the compositionality of the implementation
relation:
\begin{lemma} \label {oeifuygrui}
$\comp_1\models\cgte_1$ and $\comp_2\models\cgte_2$ together 
imply \mbox{$\comp_1\times\comp_2 \;\models\; \cgte_1 \product \cgte_2$.}
\end{lemma}
\IEEEproof
The assumption of the lemma means that
$\comp_i\subseteq\maxcomp{\cgte_i}$, for $i=1,2$. Since the two
contracts are in canonical form, we have $\maxcomp{\cgte_i}=\prom_i$
and the result follows directly from Definition \ref{def:composition_restr}.
\endIEEEproof
The following lemma relates greatest lower bound and parallel
composition, it relies on the fact that we work with contracts in
canonical form:
\begin{lemma} \label {wrotheortui}
For any two contracts, 
$\cgte_1\meet\cgte_2 \cgref \cgte_1\para\cgte_2$.
\end{lemma} 
\IEEEproof
Both sides of this relation possess identical promises. Thus the only
thing to prove relates to the assumptions thereof.  From
Definition~\ref{def:composition_restr} and
Theorem~\ref{th:bounds_restr}, the assumption of $\cgte_1\meet\cgte_2$
is equal to $\assp_1\cup\assp_2$, whereas the assumption of
$\cgte_1\para\cgte_2$ is equal to $(\assp_1 \intersect \assp_2) \union
\complement{(\prom_1 \intersect \prom_2)}$. Since the two contracts
are in canonical form, we have $\complement{\prom_i}\subseteq\assp_i,
i=1,2$, and thus $\complement{(\prom_1\cap\prom_2)}
=\complement{\prom_1}\cup\complement{\prom_2}
\subseteq\assp_1\cup\assp_2$. Therefore, the assumption of
$\cgte_1\para\cgte_2$ is contained in $\assp_1\cup\assp_2$, which is
the assumption of $\cgte_1\meet\cgte_2$. This proves the lemma.
\endIEEEproof
\begin{example}{Running example: compositional reasoning}
\label {;erfouihe;uio}
In Example~\ref{e;rouifhe}, we have shown how to combine the two
nominal and exception viewpoints, for the overall system of
Figure~\ref{erpfiuerhpifu}. The system further decomposes into a
control and monitoring unit. We would like to associate contracts to
each of these components, for each of their viewpoint.
Composing these contracts, we should recover the system's overall
contract.

Since the system is the parallel composition of control and
monitoring units, we may reasonably expect that the parallel
composition of contracts, for each of these components, should be
used. However, we are also combining viewpoints for these two
components and this sould be performed by the
greatest lower bound. So, which is the correct answer?
%
The new notion of contract
\emph{fusion} we shall introduce in the following section will
provided the adequate answer. Prior to introducing this notion, we need
to investigate what it means to eliminate ports in contracts.
\end{example}

\subsubsection{\textbf{Eliminating ports in contracts}}
Elimination in contracts requires handling assumptions and promises
differently.
\begin{definition} [Elimination]
\label {erotijherio}
Let $\cgte=(\assp,\prom)$ be a contract 
and let $p$ be any port. Define the \emph{elimination of $p$
in $\cgte$ by:} 
\beqq 
\elim{\cgte}{p} &=& (\forall p \,\assp, \exists p \,
\prom) 
\eeqq
where $\assp$ and $\prom$ are seen as predicates.
\end{definition}
Note that we do not require that $p$ be a port of $\cgte$.
Definition~\ref{erotijherio} is motivated by the following lemma:
\begin{lemma}\label{erptihpui}
We have $\cgte\cgref\elim{\cgte}{p}$. Furthermore, let $\comp$ be an
implementation such that $\comp\models\cgte$ and $p$ is not a port of
$\comp$.  Then, $\comp\models\elim{\cgte}{p}$.
\end{lemma}
\IEEEproof
By definition, $\comp\models\cgte$ implies $\comp\cap\assp\subseteq\prom$.
Eliminating $p$, with
$\forall$ on the left hand side and $\exists$ on the right hand side,
yields $[\forall p\,(\comp\cap\assp)]\subseteq[\exists p\,\prom]$ and
the lemma follows from the fact that $\forall p\,(\comp\cap\assp)
=\comp\cap(\forall p\,\assp)$ if $p$ is not a port of $\comp$.
\endIEEEproof
The following lemma relates elimination and greatest lower bounds:
%
\begin{lemma}  \label {erptfuhpeioru}
For any two contracts $\cgte_1$ and $\cgte_2$ and any port $p$, we have:
\beq
\elim{\cgte_1 \meet \cgte_2}{p} 
&\cgref&
\elim{\cgte_1}{p} \meet \elim{\cgte_2}{p} 
\label {we;rofh;sro}
\\
\elim{\cgte_1 \para \cgte_2}{p} 
&\cgref&
\elim{\cgte_1}{p} \para \elim{\cgte_2}{p} 
\label {erptheruih}
\eeq
\end{lemma}
\IEEEproof
We have $\forall p\,(\assp_1\cup\assp_2)\supseteq(\forall
p\,\assp_1\cup\forall p\,\assp_2)$ and $\exists
p\,(\prom_1\cap\prom_2)\subseteq$ \mbox{$(\exists
p\,\prom_1\cap\exists p\,\prom_2)$,} which proves (\ref{we;rofh;sro})
as well as the promise part of (\ref{erptheruih}). Regarding the
assumption part of (\ref{erptheruih}), we need to prove 
\beq
\assp_{\elim{\cgte_1 \para \cgte_2}{p}} &\supseteq& 
\assp_{(\elim{\cgte_1}{p} \para
\elim{\cgte_2}{p})} 
\label {erotihepro}
\eeq
where
\beqq
\assp_{\elim{\cgte_1 \para \cgte_2}{p}} &=& \forall p\,\bigl(
(\assp_1\cap\assp_2)\cup\complement{(\prom_1\cap\prom_2)}
\bigr)
\\
\assp_{(\elim{\cgte_1}{p} \para
\elim{\cgte_2}{p})} &=& (\forall p\,
\assp_1\cap\forall p\,
\assp_2)\cup\complement{(\exists p\,\prom_1\cap\exists p\,\prom_2)}
\eeqq
We have 
%
%
$\forall p\,((\assp_1\cap\assp_2)\cup\complement{(\prom_1\cap\prom_2)}) 
\, \supseteq \,
(\forall p\,(\assp_1\cap\assp_2)) 
\cup
(\forall p\,\complement{(\prom_1\cap\prom_2)}) 
\, = \,
(\forall p\,\assp_1\cap
\forall p\,\assp_2) 
\cup
\complement{(\exists p\,(\prom_1\cap\prom_2))}
\, \supseteq \,
(\forall p\,
\assp_1\cap\forall p\,
\assp_2)\cup\complement{(\exists p\,\prom_1\cap\exists p\,\prom_2)}$.
Which proves (\ref{erotihepro}) and the lemma.
\endIEEEproof
Elimination trivially extends to finite sets of ports, we denote it by 
$\elim{\cgte}{P}$, where $P$ is the considered set of ports.

\section{Set of contracts associated to a rich component}
\label {epfguioioe}
We are now ready to address the case of synchronizing viewpoints when
local ports are shared between viewpoints. More precisely, we shall
formally define what it means to consider a set of contracts
associated to a same rich component.

\begin{definition}[Fusion] \label {sr;otihjeriot}
Let $(\cgte_i)_{i\in I}$ be a finite set of contracts and $Q$ a finite
set of ports. We define \emph{the fusion of $(\cgte_i)_{i\in I}$ with
respect to $Q$} by
\beq 
\fusion{(\cgte_i)_{i\in I}}{Q} &=& \bigmeet_{J\subseteq I}
\bigelim{\,
\|_{j\in J} \cgte_j
\,}{Q}
\label {e;'rotij;i}
\eeq
where $J$ ranges over the set of all subsets of $I$.
\end{definition}
The following particular cases of Definition \ref{sr;otihjeriot} are
of interest:
\begin{lemma} \
\begin{enumerate}
\item \label {;erofuihe}
When $Q=\emptyset$, the fusion reduces to the greatest lower
bound: 
\beq 
\fusion{(\cgte_i)_{i\in I}}{\emptyset} &=& \meet_{i\in I}
\cgte_i
\label {rpotuihleui}
\eeq
In particular, $\comp\models\fusion{(\cgte_i)_{i\in I}}{\emptyset}$ implies 
$\comp\models\cgte_i$ for each ${i\in I}$.

\item \label {perfuohu}
Assume that, for $i=1,2$:
\beq
\assp_i&\supseteq&\prom_1\cap\prom_2
\label {oeirtuheoriu}
\eeq
holds. Then:
\beq
\fusion{(\cgte_i)_{i\in \{1,2\}}}{\emptyset} &= &
\cgte_1\para\cgte_2 
\label {er;otijhpeotie}
\eeq
\item \label {edfguioeriofg}
Assume that, for $i=1,2$:
\beq
\forall Q\, (\assp_i\cup\complement{\prom}) &\supseteq&
\forall Q\, (\assp_1\cup\assp_2)
\label {w;otijpeio}
\eeq
holds, where $\cgte_1\para\cgte_2=(\assp,\prom)$. Then:
\beq
\fusion{(\cgte_i)_{i\in \{1,2\}}}{Q} &=& \elim{\cgte_1\para\cgte_2}{Q} 
\label {etih;ui}
\eeq
\end{enumerate}
\end{lemma}
Condition (\ref{oeirtuheoriu}) expresses that each rich component is a
valid environment for the other rich components; in other words, the
two contracts are attached to two rich components that together
constitute a valid closed system.
Condition (\ref{w;otijpeio}) expresses that the restriction to $Q$
of each component is a valid environment for the restriction to $Q$
of the other component. This situation corresponds to two rich components
interacting through ports belonging to $Q$, which are
subsequently hidden from outside.
\IEEEproof
We successively prove the three statements.

Statement \ref{;erofuihe} results immediately from Lemma \ref{wrotheortui}.

To prove (\ref{er;otijhpeotie}) in Statement \ref{perfuohu}, note that the two
expressions only differ by their assumptions, since the promises of
greatest lower bound and parallel composition are identical. For the assumptions, let
$\cgte_1\para\cgte_2=(\assp,\prom)$ and
$\cgte_1\meet\cgte_2=(\assp',\prom)$. We have
$\prom=\prom_1\cap\prom_2$, 
$
\assp = (\assp_1\cap\assp_2)\cup\complement{\prom}
$, and $
\assp' = (\assp_1\cup\assp_2)
$. 
From (\ref{oeirtuheoriu}) we get
$\complement{\prom}\supseteq\complement{(\assp_1\cap\assp_2)}$. Therefore,
$\assp=(\assp_1\cap\assp_2)\cup
\complement{\prom}\supseteq(\assp_1\cap\assp_2)\cup
\complement{(\assp_1\cap\assp_2)}=\runs\supseteq\assp'$.

Regarding Statement \ref{edfguioeriofg},
(\ref{w;otijpeio}) implies 
$
\forall Q\, ((\assp_1\cap\assp_2)\cup\complement{\prom})
\supseteq
\forall Q\, (\assp_1\cup\assp_2)
$. Whence (\ref{etih;ui}) follows.
\endIEEEproof
The lesson is that fusion boils down to parallel composition for
contracts attached to two different sub-components of a same compound
component, whereas contracts attached to a same component and
involving the same set of ports fuse via the operation of greatest
lower bound.  The general case lies in between and is given by formula
(\ref{e;'rotij;i}).  
Finally, the various relations that we have established between
greatest lower bound, parallel composition, and elimination, allows us
to simplify the actual evaluation of the fusion in
general. Corresponding heuristics to guide this remain to be
developed.

%
Definition \ref{epfiueipu} for rich components can now be completed.
\begin{definition} [Rich Component, completed] \label {pweifugbepruif}
Let $\richcomp = \bigl(\rcname, \setsofcontracts,
\optimplementation\bigr)$ be a rich component.  Say that
\beqq
\optimplementation\models\setsofcontracts &\mbox{iff}&
\optimplementation\models\fusion{(\cgte_i)_{i\in I}}{Q},
\eeqq
where $I$ indexes set $\setsofcontracts$, and set $Q$ collects the
ports of $\setsofcontracts$ that are local to $\richcomp$.
\end{definition}

\begin{example}{Running example: fusion of contracts}
\label {leprifue}

We shall perform a composability study for the two contracts $\cgte$
and $\cgte'$, and then for their fusion $\fusion{\cgte,\cgte'}{}$.

\paragraph{Study of $\cgte$}
Consider the following two contracts, for the control and monitoring unit,
respectively
%
%
$\cgte_1 \, = \, (\,\neg f_1\,,\,[x=a\wedge b]\,)$ and
$\cgte_2 \, = \, (\,\neg \varphi\,,\,\,y=x\,)$, where
$\varphi = \neg a \wedge x$.
Contract $\cgte_1$ states that, if not faulty,
the control unit guarantees that $\neg\varphi$ holds, i.e., invariant
$a\vee\neg x$ holds. Contract $\cgte_2$ states that the monitoring unit
guarantees that, if not faulty, $y=x$ holds unless $\varphi$ does not
hold.  Putting these two contracts in canonical form and then
computing their fusion yields
%
%
\beqq
\cgte_1 &=& (\,\neg f_1\,,\,\neg f_1\Ra[x=a\wedge b]\,)
\nonumber \\
\cgte_2 &=& (\,\neg \varphi \,,\,\neg \varphi \Ra\,y=x\,)
\nonumber \\
\elim{\cgte_1}{x} &=& (\,\neg f_1\,,\,\true\,)
\label {eprogfuiehu}
\\
\elim{\cgte_2}{x} &=& (\,\false\,,\,\true\,)
\label {erpfgouierhpogui}
\\
\elim{\cgte_1 \para \cgte_2}{x} &=& (\,\neg f_1 \wedge P \,,\,
\neg f_1 \Ra[y=a\wedge b]\,) 
\label {erpfiubiu}
\eeqq
where $P$ is some predicate (which we don't care about), from which
we obtain, provided that $\cgte$ is put in canonical form,
\beqq
\fusion{\cgte_1,\cgte_2}{x}
= \; \elim{\cgte_1}{x} \meet \elim{\cgte_2}{x} \meet \elim{\cgte_1 \para 
\cgte_2}{x} 
\; = \;
\elim{\cgte_1 \para \cgte_2}{x} \; = \; \cgte
\label {perifguehipu}
\eeqq
%
%

\paragraph{Study of $\cgte'$}

Now, let us focus on the other contract $\cgte'$, by proposing the
following two local contracts, for the control and monitoring unit,
respectively
%
%
$\cgte'_1 \, = \, (\,\false\,,\,\true\,)$, and
$\cgte'_2 \, = \, (\,\neg f_2 \,,\,[y=x\wedge a]\,)$.
The first contract is trivial, and the second one states the invariant
promised if the monitoring unit is not faulty.
We first have 
\beq
\mbox{$\elim{\cgte'_1 }{x}=\cgte'_1 $ and 
$\elim{\cgte'_2}{x}=(\,\neg f_2 \,,\,\neg f_2 \Ra\neg\tle\,)$.}
\label {erpofguih}
\eeq
Second, $\prom'_1\cap\prom'_2\,=\,\neg f_2 \Ra[y=x\wedge a]$, whence
%
$\exists x:(\prom'_1\cap\prom'_2 ) \ = \ \neg f_2 \Ra\neg\tle$.
Next, 
$ 
(\assp'_1\cap\assp'_2)\cup\neg(\prom'_1\cap\prom'_2 )
=
\neg(\prom'_1\cap\prom'_2 )$, which equals
$\neg(\neg f_2 \Ra[y=x\wedge a])$,
whence 
\beq
\forall x:\left(
(\assp'_1\cap\assp'_2)\cup\neg(\prom'_1\cap\prom'_2 )
\right) &=& \neg(\exists x:(\prom'_1\cap\prom'_2 ))
\nonumber \\
&=& \tle\wedge\neg f_2
\label {pwerfiouwhpfuio}
\eeq
%
%
Finally, (\ref{erpofguih})--(\ref{pwerfiouwhpfuio}) together prove
that 
$
\fusion{\cgte'_1,\cgte'_2}{x} \ = \ \cgte'
$.

\paragraph{Study of $\fusion{\cgte,\cgte'}{}$}
The remarkable point is that composability works both across components,
and viewpoints/modes, i.e., we have
$
\fusion{\cgte,\cgte'}{} = \fusion{\cgte_1,\cgte_2,\cgte'_1,\cgte'_2}{x}
$.
\end{example}

\section{The asymmetric role of ports}
\label {fuohepuo}
So far we have ignored the role of ports and the corresponding
splitting of responsibilities between the implementation and its
environment, see the discussion in the introduction. 
Such a splitting of responsibilities avoids the competition
between environment and implementation in setting the value of ports and
variables.

Intuitively, an implementation can only provide promises on the value of the
ports it controls.
On ports controlled by the environment, instead, it may only declare
assumptions.
Therefore, we will distinguish between two kinds of ports for
implementations and {\contracts}: those that are \emph{controlled} and those
that are \emph{uncontrolled}.
The latter property is formalized via the following notion of
\emph{receptiveness}:
\begin{definition}[Receptiveness] \label {erpthoi}
For $\behav$ an assertion, and $P' \subseteq P$ a subset of its ports,
$\behav$ is said to be $P'$-\emph{receptive} if and only if for all
runs $\sigma'$ restricted to ports belonging to $P'$, there exists a
run in $\sigma$ of $\behav$ such that $\sigma'$ and $\sigma$ coincide
over $P'$.
\end{definition}
In words, $\behav$ accepts any history offered to the subset $P'$ of
its ports.  Note that:
\begin{lemma} \label {er;'tgijei}
If $E$ is $P'$-receptive, then so is $E\cup E\pprime$ for
any $E\pprime$ having no extra ports or variables than those of $E$.
\end{lemma}

In some cases, different viewpoints associated with a same rich
component need to interact through some common ports. This motivates
providing a scope for ports, by partitioning them into ports that are
\emph{visible} (outside the underlying component) and ports that are
\emph{local} (to the underlying component).

\begin{definition}[Profile] \label {e;ro'fguierh;of}
A \emph{profile} is a 4-tuple
$\profile=(\vis,\loc,\uncontrolled,\controlled)$, partitioning $P$ as
\[\bea{rcccl}
P &=& \vis\uplus\loc &=& \{\textrm{visible}\} \uplus \{\textrm{local}\}
\\
P &=& \uncontrolled\uplus\controlled &=&
\{\textrm{uncontrolled}\} \uplus \{\textrm{controlled}\}
\eea
\]
%
\end{definition}
We are now ready to refine our
theory of contracts by taking the asymmetric role of ports into account. 
\begin{definition}[Implementation]  \label {eotijhroei}
An \emph{implementation} is a pair $M=(\profile,\behav)$, where
$\profile=(\vis,\loc,\uncontrolled,\controlled)$ is a \emph{profile}
over a set $P$ of ports, and $\behav$ is a $\uncontrolled$-receptive
assertion over $P$.
\end{definition}
The last requirement formalizes the fact that an implementation has no
control over the values of ports set by the environment.
Implementations refine as follows:
\begin{definition}[Implementation Refinement]\label {pwfuhwpiu}
For $\comp$ and $\comp'$ two implementations, say that $\comp$ refines
$\comp'$, written $\comp \compref \comp'$, if and only if
$\profile=\profile'$ and $\behav \subseteq \behav'$.
\end{definition}
In defining parallel composition for implementations, we need to take
into account controlled ports. Each implementation is responsible for
its set of controlled ports, and, in our theory, such responsibility
should not be shared. This motivates the following definition for our
parallel composition of implementations associated with different
rich components (whence the requirement $\loc_1\cap\loc_2=\emptyset$ in
this definition):
\begin{definition} [Parallel composition of implementations]
\label {eprotuih}
Let $\comp_1$ and $\comp_2$ be two implementations such that
$\loc_1\cap\loc_2=\emptyset$. Then,  $\comp = \comp_1 \product \comp_2$ is
defined if and only if $\controlled_1 \intersect \controlled_2 =
\emptyset$.  In that case, $\behav = \behav_1 \times \behav_2$, and:
%
%
%
\begin{align*}
\vis & = \vis_1 \union \vis_2 &
\controlled & = \controlled_1 \union \controlled_2 \\
\loc & = \loc_1 \union \loc_2 &
\uncontrolled & = (\uncontrolled_1 \union \uncontrolled_2) -
(\controlled_1 \union \controlled_2)
\end{align*}
\end{definition}

\begin{theorem} 
Implementation composition is monotonic relative to implementation
refinement.
\end{theorem}
\IEEEproof
Since profiles refine via identity, this results boils down to the well known 
monotonicity w.r.t. sets of runs.
\endIEEEproof

\begin{definition} [Contract] \label {;reouho;e}
  A \emph{contract} is a triple $\cgte=(\profile,\assp,\prom)$, where
$\profile=(\vis,\loc,\uncontrolled,\controlled)$ is a profile over a
set $P$ of ports, and $\assp$ and $\prom$ are two assertions over $P$,
respectively called the \emph{assumptions} and \emph{promises} of
$\cgte$.
$\cgte$ is called \emph{consistent} if $\prom$ is
$\uncontrolled$-receptive, and \emph{compatible} if $\assp$ if
$\controlled$-receptive.
\end{definition}
As pointed out in (\ref{erotihjpui}), \emph{contracts are in canonical
form,} meaning that $\prom\supseteq\complement{\assp}$. If this is not
the case, we simply replace $\prom$ by its most permissive version
$\prom\cup\complement{\assp}$, which cannot per se break consistency, 
thanks to Lemma \ref{er;'tgijei}.
The sets $\assp$ and $\prom$ are not required to be receptive.
However, if $\prom$ is not $\uncontrolled$-receptive, then the promises
constrain the uncontrolled ports of the {\contract}.
This is against our policy of separation of responsibilities, since we
stated that uncontrolled ports should remain entirely under the
responsibility of the environment.
Corresponding {\contracts} are therefore called \emph{inconsistent.}

\begin{definition} [Satisfaction] \label {;oriuihtoe}
  An implementation $\comp$ \emph{satisfies} {\contract} $\cgte$, written
  \mbox{$\comp\models \cgte$}, iff 
$\profile_\comp = \profile_\cgte$ and $\behav_\comp \subseteq \prom_\cgte$.
\end{definition}
By Lemma \ref{er;'tgijei}, if {\contract} $\cgte$ is consistent, then  
$\comp_\cgte=\prom_\cgte\union\complement{\assp_\cgte}$ is still
the \emph{maximal implementation} satisfying $\cgte$.
We now turn to the relation of dominance and its consequences.

\begin{definition} [Contract Dominance] 
\label {prituhpeuirth}
A {\contract} $\cgte = (\profile, \assp, \prom)$ dominates
a {\contract} $\cgte' = (\profile, \assp', \prom')$,
written $\cgte \cgref \cgte'$, if and only if
$\profile = \profile'$,
$\assp \supseteq \assp' $, and 
$\prom \subseteq \prom'$.
\end{definition}

\begin{theorem} [Boolean algebra of contracts] \label {woduygwdvuy}
Let $\cgte_1 = (\profile_1, \assp_1, \prom_1)$ and
$\cgte_2 = (\profile_2, \assp_2, \prom_2)$ be
{\contracts} such that $\profile_1=\profile_2 = \profile$.
Then $\cgte = $
\mbox{$
(\profile, \assp_1 \union \assp_2,
\prom_1 \intersect \prom_2)$} is the greatest
lower bound of $\cgte_1$ and $\cgte_2$, written $\cgte = \cgte_1 \meet
\cgte_2$.
Similarly, $\cgte' = (\profile, \assp_1 \intersect
\assp_2, \prom_1 \union \prom_2)$ is the least upper bound of
$\cgte_1$ and $\cgte_2$, written $\cgte = \cgte_1 \meet \cgte_2$.
Finally, the complement of 
$\cgte = (\profile, \assp, \prom)$ is $\complement{\cgte} =
(\profile, \complement{\assp}, \complement{\prom})$.
\end{theorem}
\IEEEproof
This is a direct consequence of Theorem \ref{th:bounds_restr}
\endIEEEproof

Finally, it remains to define the parallel composition of contracts.
Having done this we can directly borrow the definition
\ref{sr;otihjeriot} of fusion, for contracts enhanced with profiles.
\begin{definition} [Parallel composition of contracts]
\label {weouirhpeouih}
Let $\cgte_1 = (\profile_1, \assp_1, \prom_1)$ and $\cgte_2 = (\profile_2,
\assp_2, \prom_2)$ be {\contracts}.
The parallel composition, or product, of $\cgte_1$ and $\cgte_2$, written
$\cgte = \cgte_1 \product \cgte_2$, is defined if and only if $\controlled_1
\intersect \controlled_2 = \emptyset$, and in that case is the {\contract}
$\cgte = (\profile, \assp, \prom)$ defined by:
\begin{eqnarray*}\bea{rcl}
\vis & = & \vis_1 \union \vis_2, \\
\loc & = & (\loc_1 \union \loc_2)-(\vis_1 \union \vis_2), \\
\controlled & = & \controlled_1 \union \controlled_2, \\
\uncontrolled & = & (\uncontrolled_1 \union \uncontrolled_2) -
(\controlled_1 \union \controlled_2), \\
\assp & = & (\assp_1 \cap \assp_2) \union \complement{(\prom_1
\cap \prom_2)}, \\
\prom & = & \prom_1 \cap \prom_2.
\eea
\end{eqnarray*}

\end{definition}
Unlike Definition~\ref{eprotuih}, we do not require here that 
$\loc_1 \cap \loc_2=\emptyset$. The reason is that 
we wish to encompass the composition of different viewpoints
attached to a same rich component. (For contracts attached to different
rich components, however, we do have $\loc_1 \cap \loc_2=\emptyset$.)

With parallel composition, we can formalize the notion of contract
compatibility.
Recall that a contract is \emph{compatible} whenever $\assp$ is
$\controlled$-receptive.
If not, then there exists a sequence of values on the controlled ports that
are refused by all acceptable environments.
However, by our definition of satisfaction, implementations are allowed to
output such sequence.
Unreceptiveness, in this case, implies that a hypothetical environment that
wished to prevent a violation of the assumptions should actually prevent the
behavior altogether, something it cannot do since the port is controlled by
the {\contract}.
Therefore, unreceptive assumptions denote the existence of an
incompatibility internal to the contract, that cannot be avoided by any
environment.
This justifies the following definition.
\begin{definition}[Compatibility] Two contracts $\cgte_1$ and $\cgte_2$ are
  \emph{compatible} if and only the assumption of their parallel composition
  is receptive with resepct to the controlled ports.
\end{definition}
Assumptions may become unreceptive as a result of a parallel composition
even if they are not so individually.
This is because the set of controlled ports after a composition is strictly
larger than before the composition.
In particular, ports that were uncontrolled may become controlled, because
they are controlled by the other {\contract}.



Note that consistency and compatibility may not be preserved by Boolean
operations.
For example, one obtains an inconsistent {\contract} when taking the greatest
lower bound of two {\contracts}, one of which promises that certain behaviors
will never occur in response to a certain input, while the other
promises that the remaining behaviors will not occur in response to the same
input.
Both {\contracts} have legal responses to the input,
but their intersection is empty,
thus making the combination
unreceptive.
In this case, inconsistency is due to two {\contracts} making inconsistent
promises.



%


\section{Addressing Multiple Viewpoints}
\label {er;ofuierhopuif}
An important question is: can our abstract notion of ``assertion''
encompass the different functional and non-functional
viewpoints of system design?
Since assertions are just sets of runs, we can, in particular,
accomodate hybrid automata following~\cite{Henzinger96}. 
So seemingly, we can in particular support 
functional, timeliness, safety, as
these can be modeled by specific subclasses of hybrid automata.

A closer investigation reveals that we need to deal with classes of
models that are stable under parallel composition (defined by
intersection), union, and complement.
Taking complements is a delicate issue: hybrid automata are not closed
under complementation; in fact, no model class is closed under
complementation beyond deterministic automata.
To account for this fact, various countermeasures can be considered.

First, the designer has the choice to specify either $\behav$ or its
complement $\neg\behav$ (e.g., by considering observers).
However, the parallel composition of contracts requires manipulating
both $\behav$ and its complement $\neg\behav$, which is the embarrasing
case. 
To get compact formulas, our theory was developed using canonical
forms for contracts, systematically. 
Not enforcing canonical forms provides room for flexibility in the
representation of contracts, which can be used to avoid manipulating
both $\behav$ and $\neg\behav$ at the same time.
A second idea is to redefine an assertion as a \emph{pair}
$(\behav,\bar\behav)$, where $\bar\behav$ is an approximate complement
of $\behav$, e.g., involving some abstraction.
In doing so, one of the two characteristic properties of
complements, namely $\behav\cap\bar\behav=\emptyset$ or
$\behav\cup\bar\behav=\top$, do not hold.
%
%
However, either necessary of sufficient conditions for contract
dominance can be given. The above techniques are the subject of
ongoing work and will be reported elsewhere.

\section{Related Work}\label{sec:related}
The notion of contract derives from the theory of
abstract data types, first suggested by Meyer in the context of the
programming language Eiffel~\cite{Meyer92IEEEC}.
In his work, Meyer introduces \emph{preconditions} and \emph{postconditions}
as assertions for the methods of a class, and
\emph{invariants} for the class itself.
Preconditions correspond to the assumptions under which the method operates,
while postconditions express the promises at 
method termination, provided that the assumptions are satisfied.
Invariants must be true at all states 
of the class regardless of any assumption.
To guarantee safe substitutability, a subclass is only allowed to weaken the
assumptions and to strengthen the promises.

Similar ideas were 
in fact, already present in earlier work by Dill, although phrased in less
explicit terms~\cite{Dill89ThesisACM}.
Dill proposes an asynchronous model based on sets of sequences
and parallel composition (trace structures).
Behaviors (traces) can be either accepted as \emph{successes}, or rejected
as \emph{failures}.
The failures, which are still possible behaviors of the system, correspond
to unacceptable inputs from the environment, and are therefore the
complement of the preconditions.
Safe substitutability is expressed as 
trace containment between the successes and failures of the
specification and the implementation.
Wolf later extended the same technique to a discrete synchronous
model~\cite{Wolf95Thesis}.
More recently, De Alfaro and Henzinger have proposed Interface Automata
which are similar to synchronous trace structures, where failures are
implicitly all the traces that are not accepted by an automaton representing
the component~\cite{DeAlfaroHenzingerFSE2001}.
Composition is defined on automata, rather than on traces, and requires a
procedure to restrict the state space that is equivalent to the process
called autofailure manifestation of Dill and Wolf.
A more general approach along the lines proposed by Dill and Wolf is the
work by Negulescu with Process Spaces~\cite{Negulescu98Thesis}, and by
Passerone with Agent Algebra~\cite{Passerone04Thesis}, both of which extend
the algebraic approach to generic behaviors introduced by
Burch~\cite{Burch92Thesis}.

Our notion of contract supports \emph{speculative design} in which distributed
teams develop partial designs concurrently and synchronize by relying
on the notions of rich component~\cite{Damm05} and associated contracts.
We define assumptions and promises in terms of behaviors, and use
parallel composition as the main operator for decomposing a design.
This choice is justified by the reactive nature of embedded software,
and by the increasing use of component models that support structured
concurrency.  
We developed our theory on the basis of assertions, i.e., 
languages of generic ``runs''.  
To achieve the generality of a (mathematical) metamodel we
have complemented this by developing 
a concrete model for such assertions, that encompasses
the different viewpoints of the design~\cite{speeds}.

In our approach, behaviors are decomponsed into
assumptions and promises, as in Process Spaces, a representation that
is more intuitive than, albeit equivalent to, the one based on the
successes and failures of asynchronous trace structures.  Unlike
Process Spaces, however, we explicitly consider inputs and outputs,
which we generalize to the concept of controlled and uncontrolled
signals.  This distinction is essential in our framework to determine
the exact role and responsibilities of users and suppliers of
components.  
This is concretized in our framework by a notion of
compatibility which depends critically on the particular partition of
the signals into inputs and outputs.  
We also extend the use of
receptiveness of asynchronous trace structures, which is absent in
Process Spaces, to define formally the condition of compatibility of
components for open systems.

Our refinement relation between contracts, which we call \emph{dominance} to
distinguish it from refinement between implementations of the contracts,
follows the usual scheme of weakening the assumption and strengthening the
guarantees.
The order induces boolean operators of conjunction and disjunction, which
resembles those of asynchronous trace structures and Process Spaces.
In addition, we also define a new \emph{fusion} operator that combines the
operation of composition and conjunction for a set of contracts.
This operator is introduced to make it easier for the user to express the
interaction between contracts related to different viewpoints of a
component.

The model that we present in this paper is based on execution traces, and is
therefore inherently limited to representing linear time properties.
The branching structure of a process whose semantics is 
expressed in our model is thus abstracted, and the exact state in which
non-deterministic choices are taken is lost.
Despite this, the equivalence relation that is induced by our notion of
dominance between contracts is more distinguishing than the traditional
trace containment used when executions are not represented as pairs
(assumptions, promises).
This was already observed by Dill, with the classic example of the vending
machine~\cite{Dill89ThesisACM}, see also Brookes et al.\ on 
refusal sets~\cite{BrookesHoareRoscoe84JACM}.
There, every accepted sequence of actions is complemented by the set of
possible \emph{refusals}, i.e., by the set of actions that may not be
accepted after executing that particular sequence.
Equivalence is then defined as equality of sequences with their refusal
sets.
Under these definitions, it is shown that the resulting equivalence is
stronger than trace equivalence (equality of trace sets), but weaker than
observation equivalence~\cite{Engelfriet,Brookes83ICALP}.
A precise characterization of the relationships with our
model, in particular with regard to the notion of composition, is deferred
to future work.


\section{Conclusion}
We have presented mathematical foundations for the contract-based
model developed in the framework of the SPEEDS project.  Our generic
mathematical model of contract supports ``speculative design''.  This
is achieved by focusing on behaviors, by supporting the notion of
rich component where diverse (functional and non-functional) aspects
of the system can be considered and combined, by representing rich
components via their set of associated contracts, and by formalizing
the whole process of component composition through the general
mechanism of contract fusion. These foundations support the
Heterogeneous Rich Component (HRC) metamodel under development in
SPEEDS~\cite{speeds}.
Future work includes the development of effective algorithms to handle
contracts, coping with the problems raised by complementation.

\textbf{Acknowledgements:}~This research has been developed in the
framework of the SPEEDS integrated European project number 033471. We
would like to thank all SPEEDS project participants for the fruitful
discussions we had with them, and for the suggestions they made to
improve the research report.

\bibliographystyle{IEEEtran}
\bibliography{IEEEabrv,speeds}

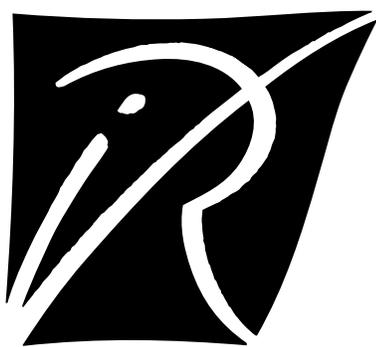
\end{document}